\documentclass[reviewcopy]{elsart}

\usepackage{graphicx}
\usepackage{amsfonts,amsmath}

\newcommand{\Tr}{{\textrm{Tr}}}

\newcommand{\ket}[1]{|#1\rangle}
\newcommand{\bra}[1]{\langle #1|}
\begin{document}
\begin{frontmatter}
\title{Entanglement-induced geometric phase of quantum states}
\author{Erik Sj\"oqvist},
\ead{eriks@kvac.uu.se}
\address{Department of Quantum Chemistry, Uppsala University,
Box 518, Se-751 20 Uppsala, Sweden}
\date{\today}
\begin{abstract}
The concept of relative state is used to introduce geometric phases
that originate from correlations in states of composite quantum systems.
In particular, we identify an entanglement-induced geometric
phase in terms of a weighted average of geometric phase factors
associated with a decomposition that define the entanglement of
formation. An explicit procedure to calculate the
entanglement-induced geometric phase for qubit pairs is put
forward. We illustrate it for maximally entangled mixed states (MEMS)
of two qubits.
\end{abstract}
\end{frontmatter}
\maketitle
A quantum system may pick up a geometric phase when it evolves
along a path in its state space \cite{berry84,aharonov87}. This
phase becomes additive for product states of composite systems
since the uncorrelated subsystems pick up independent geometric phase
factors. However, in the presence of quantum correlations the situation
becomes less clear in that there is no unique assignment of phase
factors to each subsystem in this case. This is related to the
existence of several inequivalent forms of mixed state geometric phases
\cite{uhlmann86,sjoqvist00,chaturvedi04,marzlin04,slater02,ericsson03,rezakhani06},
leading to different phase assignments when these forms are applied to the non-pure marginal
states of the correlated subsystems. The effort to further clarify the role of
geometric phases in composite systems has triggered attempts to find
alternative geometric phase concepts and geometric phase-like structures
that may capture explicitly the correlation structure of quantum states
\cite{wootters02,levay05,williamson07,sjoqvist09}.

Recently, such an alternative concept was proposed \cite{sjoqvist09} based
on Everett's relative state \cite{everett57}. This geometric
phase is induced by quantum entanglement, if the full state is pure. On
the other hand, classical correlations and quantum entanglement can coexist
in mixed quantum states, which implies that the forms of mixed state geometric
phases in \cite{uhlmann86,sjoqvist00}, applied to the path of
relative states, may contain contributions from both these types of
correlations. Here, we show that one can take advantage of the decomposition
freedom of mixed quantum ensembles \cite{hughston93} to develop another
notion of relative state based geometric phase that is entanglement-induced
in the sense that it can be non-zero only for inseparable (entangled)
states. We find an explicit procedure to calculate this entanglement-induced
geometric phase in the case of qubit systems.

Consider a mixed state of a composite quantum system described by
a normalized density operator $\varrho$ on the bipartite tensor product
decomposition $\mathcal{H}_A \otimes \mathcal{H}_B$ corresponding to
two physical subsystems $\mathcal{S}_A$ and $\mathcal{S}_B$. This
state may contain both entanglement and classical correlations. The operator
$\varrho(\phi) = \bra{\phi} \varrho \ket{\phi}$, $\ket{\phi} \in
\mathcal{H}_A$, on $\mathcal{H}_B$ is positive with
$\Tr \varrho(\phi) \leq 1$. We take it to be the state of $\mathcal{S}_B$
relative $\phi$. For a path $L: [0,1] \ni s\mapsto \phi_s$ in projective
Hilbert space $\mathcal{P} (\mathcal{H}_A)$, $\varrho(\phi_s)$
traces out a path in the space $\mathcal{T} (\mathcal{H}_B)$ of positive
linear operators acting on $\mathcal{H}_B$. Associated with
this path in $\mathcal{T} (\mathcal{H}_B)$, which is induced by the path $L$ in
$\mathcal{P} (\mathcal{H}_A)$, we wish to define geometric phases that
reflect the correlation structure of $\varrho$.

Assume that the pure states $\{ \psi_k \}_k$ and probabilities
$\{ p_k \}_k$ constitute a decomposition of $\varrho$. Let
$\{ \ket{\psi_k} \}_k$ be a set of ``subnormalized'' vectors
(i.e., $p_k=\bra{\psi_k}\psi_k \rangle$) representing these
states in $\mathcal{H}_A \otimes \mathcal{H}_B$. We may write
$\varrho = \sum_k \ket{\psi_k} \bra{\psi_k}$.
A path $L:[0,1] \ni s\mapsto \phi_s$ in projective Hilbert space
$\mathcal{P} (\mathcal{H}_A)$ defines a
geometric phase distribution $\{ \gamma_{\psi_k(\phi)}, \bra{\psi_k}
\psi_k \rangle \}_k$ with each pure state geometric phase
$\gamma_{\psi_k(\phi)}$ given by
\begin{eqnarray}
\gamma_{\psi_k (\phi)} & = & \arg \langle \psi_k (\phi_0)
\ket{\psi_k (\phi_1)} - {\textrm{Im}} \int_0^1
\frac{\langle \psi_k (\phi_s) \ket{\psi_k (\dot{\phi}_s)}}
{\langle \psi_k (\phi_s) \ket{\psi_k (\phi_s)}} ds
\nonumber \\
 & = & \arg \bra{\phi_1} \rho_{A;k} \ket{\phi_0} +
{\textrm{Im}} \int_0^1
\frac{\bra{\phi_s} \rho_{A;k} \ket{\dot{\phi}_s}}
{\bra{\phi_s} \rho_{A;k} \ket{\phi_s}} ds ,
\label{eq:relgp1}
\end{eqnarray}
where $\rho_{A;k} = \Tr_B \ket{\psi_k} \bra{\psi_k}$. Equation
(\ref{eq:relgp1}) reduces to that of \cite{sjoqvist09} when
$L$ is a loop, i.e, in the case of cyclic evolution. Note, in
particular, the important properties $\gamma_{\psi_k (\phi)} = 0$ if
$\psi_k$ is a product state and $\gamma_{\psi_k (\phi)} = -\gamma_{\phi}$
if $\psi_k$ is maximally entangled\footnote{The minus sign originates
from the antilinear nature of the relative state map $\phi \mapsto
\psi_k (\phi)$ \cite{kurucz01}.}.
Following \cite{marzlin04}, the first moment of the phase
distribution function $P(\eta)=\sum_k \bra{\psi_k}\psi_k \rangle
\delta \left( e^{i\eta} - e^{i\gamma_{\psi_k(\phi)}} \right)$
defines the geometric phase $\Gamma$ as
\begin{eqnarray}
e^{i\Gamma} =
\Phi \left( \left\langle e^{i\eta} \right\rangle_P \right) =
\Phi \left( \sum_k \bra{\psi_k}\psi_k \rangle
e^{i\gamma_{\psi_k (\phi)}} \right)
\end{eqnarray}
with $\Phi (z) = z/|z|$ for any non-zero complex number $z$. The spectral
decomposition $\{ \ket{\varphi_k} \}_k$ of $\varrho$ naturally defines a
geometric phase $\Gamma_{\varrho (\phi)}$ that is induced by the overall
correlations (classical or quantum) in the sense that it may be nontrivial
only for non-product states, i.e., if $\varrho \neq \Tr_B \varrho \otimes
\Tr_A \varrho$. We now wish to find another decomposition that can identify
the entanglement-induced contribution $\Gamma_{\varrho (\phi)}^E$ to
$\Gamma_{\varrho (\phi)}$.

We base our choice of preferred decomposition on entanglement of
formation $E$ \cite{bennett96}, which for a pure bipartite state
$\Psi$ equals the von Neumann entropy of the reduced state of
any of the subsystems, i.e., $E(\Psi)=-\Tr (\rho_A
\log_2 \rho_A) = -\Tr (\rho_B \log_2 \rho_B)$ with $\rho_A =
\Tr_B \ket{\Psi} \bra{\Psi}$ and $\rho_B =
\Tr_A \ket{\Psi} \bra{\Psi}$. For a mixed state $\varrho$,
$E$ is defined as \cite{bennett96}
$E (\varrho) = \min \sum_k \bra{\psi_k} \psi_k \rangle E(\psi_k)$,
where the minimum is taken over all decompositions $\{ \psi_k \}_k$
of $\varrho$. Let $\{ \zeta_k \}_k$ be such an entanglement-minimizing
decomposition. We take $\{ \zeta_k \}_k$ to be the preferred decomposition
and define the entanglement-induced geometric phase
$\Gamma_{\varrho (\phi)}^E$ as
\begin{eqnarray}
e^{i\Gamma_{\varrho (\phi)}^E} =
\Phi \left( \sum_{k=1}^n \bra{\zeta_k} \zeta_k \rangle
e^{i\gamma_{\zeta_k (\phi)}} \right) .
\label{eq:relgp2}
\end{eqnarray}
Note that both $\Gamma_{\varrho (\phi)}^E$ and $\Gamma_{\varrho (\phi)}$
reduce to Eq. (\ref{eq:relgp1}) in the pure state limit. In
other words, there are no contributions to $\Gamma_{\varrho (\phi)}$
from classical correlations if the full state is pure. Furthermore,
we see that $e^{i\Gamma_{\varrho (\phi)}^E} = 1$ for separable
states.

For mixed qubit-pair states, concurrence $C(\varrho)$ determines uniquely
$E(\varrho)$ and the corresponding entanglement-minimizing decomposition
can be found \cite{wootters98}. Explicitly, $C(\varrho) = \max \{ 0,\sqrt{\lambda_1} -
\sqrt{\lambda_2} - \sqrt{\lambda_3} - \sqrt{\lambda_4} \}$, where
$\lambda_k$ are the eigenvalues in decreasing order of
$\varrho \, \sigma_y \otimes \sigma_y
\varrho^{\ast} \sigma_y \otimes \sigma_y$ (``$\ast$'' stands for
complex conjugation in the computational basis). This quantity
ranges from $C(\varrho)=0$ (separable states) and $C(\varrho)=1$
(pure Bell states). In the qubit case, there exists an
entanglement-minimizing decomposition $\{ \zeta_k \}_{k=1}^n$
with optimal cardinality $n \leq 4$ such that $C(\zeta_k)=C(\varrho)$,
$\forall k$ \cite{wootters98}. We take
such an ``optimal'' decomposition to be the preferred one
for $\Gamma_{\varrho (\phi)}^E$.

To calculate $\Gamma_{\varrho (\phi)}^E$ for inseparable qubit-pair
states, we use the Wootters procedure \cite{wootters98} to find
$\{ \zeta_k \}_k$. An important element in this construction is
the existence of a subnormalized ``intermediate'' decomposition
$\{ \ket{y_k} \}_k$ of
$\varrho$, which is such that $\bra{y_1} \widetilde{y}_1 \rangle =
\sqrt{\lambda_1}$ and $\bra{y_k} \widetilde{y}_k \rangle =
-\sqrt{\lambda_k}$, $k=2,\ldots n={\textrm{rank}} (\varrho)$. One
can associate a ``preconcurrence'' $c(y_k) = \frac{\bra{y_k}
\widetilde{y}_k \rangle}{\bra{y_k} y_k \rangle}$, $k=1,\ldots,n$, to
each $y_k$. We find the averaged
preconcurrence $\langle c \rangle = \sum_k \bra{y_k} y_k \rangle
c(y_k) = C(\varrho)$, which is preserved under
orthogonal transformations $\ket{y_k} \rightarrow \ket{\zeta_k} =
\sum_l \ket{y_l} V_{lk}$. The Wootters decomposition is obtained
sequentially by letting $V$ act pairwise on the $y_k$ states with
largest and smallest preconcurrence until all preconcurrences are
equal to $C(\varrho)$. For $L: [0,1] \ni s\mapsto \phi_s$, the resulting
set of paths $\{ [0,1] \ni s\mapsto \zeta_k (\phi_s) \}_k$ is inserted
into Eq. (\ref{eq:relgp2}) and we obtain $\Gamma_{\varrho (\phi)}^E$.

\begin{figure}[htb!]
\begin{center}
\includegraphics[width=10 cm]{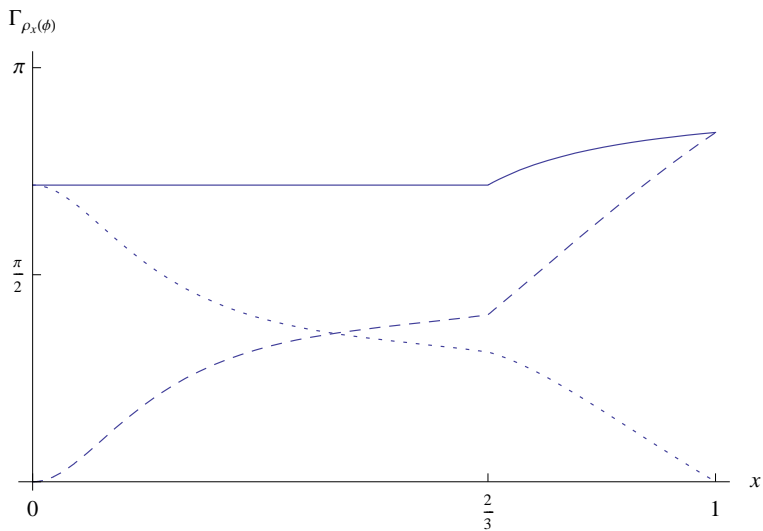}
\end{center}
\caption{\label{fig:Fig1} Correlation-induced
geometric phases $\Gamma_{\varrho_x (\phi)}$ (solid line),
$\Gamma_{\varrho_x (\phi)}^E$ (dashed line), and $\Gamma_{\varrho_x
(\phi)} - \Gamma_{\varrho_x (\phi)}^E$ (dotted line) for maximally
entangled mixed qubit states defined in Eq. (\ref{eq:munrostates})
as a function of concurrence $C(\varrho_x)=x$. The correlation-induced
mixed state geometric phase $\Gamma_{\varrho_x (\phi)}$ is the phase of
the weighted average of geometric phase factors associated with
the spectral decomposition of $\varrho_x$. We interpret
$\Gamma_{\varrho_x (\phi)}^E$ and $\Gamma_{\varrho_x (\phi)} -
\Gamma_{\varrho_x (\phi)}^E$ as the contribution to
$\Gamma_{\varrho_x (\phi)}$ from entanglement and classical
correlations, respectively. We have chosen a loop in $\mathcal{P}
(\mathcal{H}_A)$ at constant latitude $\theta = 0.45\pi$ on the
Bloch sphere. Since $\varrho_0$ is separable, $\Gamma_{\varrho_0 (\phi)}$
contains only contributions from classical correlations
($\Gamma_{\varrho_0 (\phi)}^E$ vanishes). At $x=1$ the state is pure
and therefore $\Gamma_{\varrho_1 (\phi)} =
\Gamma_{\varrho_1 (\phi)}^E$. Note that $\Gamma_{\varrho_1 (\phi)}^E =
-\gamma_{L} = \pi (1-\cos\theta) \approx 0.84 \pi$, which is
consistent with the fact that $\varrho_1$ is a Bell state. Furthermore,
$\Gamma_{\varrho_x (\phi)}^E$ has discontinuous first derivative at
$x=\frac{2}{3}$, across which
${\textrm{rank}} (\varrho_x ) =3 \rightarrow 2$.}
\end{figure}

We illustrate the above procedure for maximally entangled mixed states
(MEMS), which are two-qubit mixed states that maximize entanglement of
formation for a given purity $\Tr \varrho^2$. In the computational basis,
these states may be written as \cite{munro01}
\begin{eqnarray}
\varrho_x = \left(
\begin{array}{cccc}
g(x) & 0 & 0 & \frac{x}{2} \\
0 & 1-2g(x) & 0 & 0 \\
0 & 0 & 0 & 0 \\
\frac{x}{2} & 0 & 0 & g(x)
\end{array} \right)
\label{eq:munrostates}
\end{eqnarray}
up to local unitaries. Here, $g(x)$ is a function on $x\in [0,1]$ such that
$g(x)=\frac{1}{3}$ for $0\leq x \leq  \frac{2}{3}$ and $g(x)=\frac{x}{2}$ for
$\frac{2}{3} \leq x \leq 1$. One finds $C(\varrho_x)=\max \{ 0,x \} = x$
and subnormalized eigenvectors $\ket{\varphi^{\pm}} =
i\sqrt{\frac{p_{\pm}}{2}}(\ket{00}\pm\ket{11})$ and $\ket{\varphi^0} =
i\sqrt{p_0} \ket{01}$ with $p_{\pm}=g(x)\pm \frac{x}{2}$ and $p_0 =
1-2g(x)$. From the spectral decomposition, we find the
correlation-induced geometric phase
\begin{eqnarray}
e^{i\Gamma_{\varrho (\phi)}} =
\Phi \left( 1-2g(x) + 2g(x) e^{-i\gamma_{L}} \right) ,
\end{eqnarray}
where we have used $\gamma_{\varphi^0(\phi)}=0$ and $\gamma_{\varphi^{\pm}
(\phi)} = -\gamma_{L}$. The Wootters procedure yields the optimal
decomposition
\begin{eqnarray}
\ket{\zeta_1} & = & \frac{1}{\sqrt{2}} \left(
\sin \alpha \ket{\varphi^+} -\cos \alpha \ket{\varphi^-} +
\ket{\varphi^0} \right) ,
\nonumber \\
\ket{\zeta_2} & = & \frac{1}{\sqrt{2}} \left(
-\sin \alpha \ket{\varphi^+} +
\cos \alpha \ket{\varphi^-} + \ket{\varphi^0} \right) ,
\nonumber \\
\ket{\zeta_3} & = & \cos \alpha \ket{\varphi^+} + \sin \alpha
\ket{\varphi^-} .
\end{eqnarray}
Here, $\cos 2\alpha = \frac{f(x)}{6f^2(x)-1}$ with $f(x)=\frac{x}{2}+
\frac{1}{3}-g(x)$. Inserting into Eq. (\ref{eq:relgp2}) gives
the entanglement-induced geometric phase
\begin{eqnarray}
e^{i\Gamma_{\varrho_x (\phi)}^E} & = &
\Phi \left\{ [2-9f^2(x)] e^{i\gamma_{\zeta_1 (\phi)}} +
[2-9f^2(x)] e^{i\gamma_{\zeta_2 (\phi)}} \right.
\nonumber \\
 & & \left. + 2[1-9f^2(x)] e^{i\gamma_{\zeta_3 (\phi)}} \right\} .
\end{eqnarray}
Notice that $[1-9f^2(x)] > 0$ and $[1-9f^2(x)] = 0$ on $x\in
[0,\frac{2}{3})$ and $x\in [\frac{2}{3},1]$, respectively, i.e., the
rank of $\varrho_x$ changes from $n=3$ to $n=2$ across $x=\frac{2}{3}$.
For $\ket{\phi} = \ket{0}+z\ket{1} \in \mathcal{H}_A$ and the loop
$L_z : [0,1] \ni s \mapsto z_s$ in the complex plane corresponding
to the loop $L: [0,1] \ni s\mapsto \phi_s$ in $\mathcal{P} (\mathcal{H}_A)$,
we obtain
\begin{eqnarray}
\gamma_{\zeta_k (\phi)} =
{\textrm{Im}} \oint_{L_z} \frac{(1+\sqrt{1-x^2}) P_kdP_k^{\ast} +
(1-\sqrt{1-x^2}) Q_kdQ_k^{\ast}}{(1+\sqrt{1-x^2}) |P_k|^2 +
(1-\sqrt{1-x^2})  |Q_k|^2} .
\label{eq:purephase}
\end{eqnarray}
Here, $P_k=\mu_k+\nu_k z^{\ast}$ and $Q_k = -\nu_k + \mu_k z^{\ast}$,
where $\mu_k,\nu_k$ satisfy $\mu_k^2 + \nu_k^2 = 1$ and are determined
by $\zeta_k$ such that
\begin{eqnarray}
\nu_1 / \mu_1 & = & -\nu_2 / \mu_2 =
\sqrt{2p_v} \frac{\sqrt{p_+}\sin \alpha + \sqrt{p_-}\cos \alpha}
{p_v - \sqrt{p_+p_-}\sin 2\alpha} ,
\nonumber \\
\nu_3 & = & 0, \ \ \ \mu_3 = 1.
\end{eqnarray}
Numerical simulations of $\Gamma_{\varrho_x (\phi)}$ and
$\Gamma_{\varrho_x (\phi)}^E$ for $\arg (z)$ increasing from $0$ to $2\pi$
and $|z|=\tan \frac{\theta}{2}$ with $\theta = 0.45 \pi$ (corresponding
to a loop at constant latitude close to the equator of the Bloch sphere
$\sim \! \! \mathcal{P} (\mathcal{H}_A)$) are shown
in Fig. \ref{fig:Fig1}.  The remainder $\Gamma_{\varrho_x (\phi)} -
\Gamma_{\varrho_x (\phi)}^E$, which may be associated with the
``classical correlations'' in the states, is also shown. We see that
$\Gamma_{\varrho_x (\phi)}^E =0$ at $x=0$ and $\Gamma_{\varrho_0 (\phi)}$
contains only contribution from classical correlations. Furthermore,
$\varrho_1$ is pure and therefore $\Gamma_{\varrho_1 (\phi)}$ becomes
fully entanglement-induced; it takes the value $\Gamma_{\varrho_1
(\phi)} = \Gamma_{\varrho_1 (\phi)}^E = -\gamma_{\phi} =
\pi (1-\cos\theta) \approx 0.84 \pi$, where the
second equality follows from the definition of MEMS that entails that
$\varrho_1$ is a Bell state. At all intermediate $x$, there are
contributions to $\Gamma_{\varrho_x (\phi)}$ from both entanglement
and classical correlations. Notice, in particular, the abrupt change in
slope that occurs across the point $x=\frac{2}{3}$, which separates
the rank $n=2$ and $n=3$ regions.

The Wootters procedure may fail to give a unique decomposition in
some cases, such as when two or more of the preconcurrences
$c(y_k)$ are equal. One important example when this happens is
the class of mixed states that are diagonal in the Bell basis
$\{ \frac{1}{\sqrt{2}} ( \ket{00} \pm \ket{11} ),
\frac{1}{\sqrt{2}} ( \ket{01} \pm \ket{10} ) \}$. Such a
state is entangled iff the largest eigenvalue of the corresponding
density operator exceeds $\frac{1}{2}$. One can show that the
intermediate decomposition $\{ \ket{y_k} \}_k$ coincides with the Bell
states. Assume the largest eigenvalue belongs to $y_1$. It follows
that $c(y_1) = -c(y_2) = \ldots = -c(y_n)=1$. The non-uniqueness is now
visible: if $n = 3$ or $n = 4$, then the optimal decomposition depends
on how the state of smallest preconcurrence is chosen among
$y_2,\ldots,y_n$.

We finally demonstrate that correlation-induced geometric phases of
a mixed bipartite state $\varrho = \sum_k \ket{\psi_k} \bra{\psi_k}$ may
be implemented interferometrically as geometric phases of decomposition
dependent evolutions \cite{kult04}. Prepare the separable state
\begin{eqnarray}
\widetilde{\varrho} = \sum_k \ket{\psi_k} \bra{\psi_k} \otimes
\ket{e_k} \bra{e_k}
\end{eqnarray}
by attaching an ancilla system with Hilbert space $\mathcal{H}_e$
spanned by the orthonormal vectors $\{ \ket{e_k} \}_k$. Note that
$\varrho = \Tr_e \widetilde{\varrho}$. Let $[0,1]
\ni s \mapsto u_s$, such that $u_0 = \hat{1}_A$, be a one-parameter
family of unitary operators on $\mathcal{H}_A$ realizing the path $L: [0,1]
\ni s \mapsto \ket{\phi_s} = u_s\ket{\phi_0}$. Define another one-parameter
family of unitary operators of the form
\begin{eqnarray}
U_s^{\parallel} = \sum_k e^{i\theta_k (s)} u_s^{\dagger} \otimes \hat{1}_B
\otimes \ket{e_k} \bra{e_k}
\end{eqnarray}
acting on $\mathcal{H}_A \otimes \mathcal{H}_B \otimes \mathcal{H}_e$.
Here, all  $\theta_k$ are chosen to satisfy the parallel transport condition
$\langle \psi_k (\phi_s) \ket{\dot{\psi}_k (\phi_s)} = \langle \dot{\psi}_k
(\phi_s) \ket{\psi_k (\phi_s)}$, which amounts to
\begin{eqnarray}
\dot{\theta}_k (s) =
{\textrm{Im}} \left( \frac{\bra{\phi_s} \rho_{A;k} \ket{\dot{\phi}_s}}
{\bra{\phi_s} \rho_{A;k} \ket{\phi_s}} \right) .
\end{eqnarray}
The resulting interference pattern obtained by post-selecting
the state $\phi_0$ on the output of a Mach-Zehnder interferometer
is determined by the quantity \cite{sjoqvist00} $\Tr \bra{\phi_0} U_1^{\parallel}
\widetilde{\varrho} \ket{\phi_0} = \mathcal{V} e^{i\widetilde{\Gamma}}$,
$\mathcal{V} \geq 0$ and $\widetilde{\Gamma}$ being the visibility and
phase shift of the interference fringes. Explicitly,
\begin{eqnarray}
e^{i\widetilde{\Gamma}} =
\Phi \left( \sum_k \left| \bra{\phi_1} \rho_{A;k} \ket{\phi_0} \right|
e^{i\gamma_{\psi_k (\phi)}} \right) ,
\end{eqnarray}
i.e., correlation- and entanglement-induced
geometric phases $\widetilde{\Gamma}_{\varrho (\phi)}$
and $\widetilde{\Gamma}_{\varrho (\phi)}^E$ can be measured as shifts
in the interference fringes by choosing $\psi_k$ as the spectral
and entanglement-minimizing decompositions, respectively. Note that
the weight factors $\left| \bra{\phi_1} \rho_{A;k} \ket{\phi_0} \right|$ of
$\widetilde{\Gamma}_{\varrho (\phi)}$ and $\widetilde{\Gamma}_{\varrho (\phi)}^E$
differ from those of $\Gamma_{\varrho (\phi)}$ and $\Gamma_{\varrho (\phi)}^E$;
nevertheless $\widetilde{\Gamma}_{\varrho (\phi)}$ and $\widetilde{\Gamma}_{\varrho (\phi)}^E$
are correlation-induced in the sense that they vanish in absence of correlations
and entanglement, respectively, in the state $\varrho$. Therefore, one
may consider $\widetilde{\Gamma}_{\varrho (\phi)}$ and
$\widetilde{\Gamma}_{\varrho (\phi)}^E$ as alternative, experimentally justified,
variants of correlation- and entanglement-induced geometric phases of mixed
quantum states.

In conclusion, we have introduced geometric phases that are
induced by quantum correlations, in the sense that they may be non-zero
only for correlated quantum states. We have
identified an entanglement-induced part of this phase in terms of a
weighted average of geometric phase factors corresponding to a
decomposition that minimizes the entanglement of formation of the
state. We have demonstrated that correlation- and entanglement-induced
geometric phases may be implemented interferometrically.
\vskip 0.1 cm
Financial support from the Swedish Research Council is
acknowledged.

\end{document}